\documentclass[fleqn,usenatbib]{mnras}

\usepackage{newtxtext,newtxmath}

\usepackage[T1]{fontenc}

\DeclareRobustCommand{\VAN}[3]{#2}
\let\VANthebibliography\thebibliography
\def\thebibliography{\DeclareRobustCommand{\VAN}[3]{##3}\VANthebibliography}

\usepackage{graphicx}	
\usepackage{amsmath}	

\title[Multi-Resolution Imaging of a Radio Galaxy]{Wiggling Through the ICM: Multi-Resolution Radio Imaging of a Tailed Radio Galaxy in MACS\,J1354.6+7715}

\author[Abdul Gani et al.]{
Abdul Gani,$^{1,2}$\thanks{E-mail: agani90@gmail.com}
Roland Timmerman,$^{1,3}$
Leah K. Morabito,$^{1,3}$
Ruta Kale,$^{4}$
Satish S. Sonkamble,$^{4,5}$
\newauthor
Arpan Pal,$^{4}$
Sravani Vaddi$^{6}$
\\
$^{1}$Centre for Extragalactic Astronomy, Department of Physics, Durham University, South Road, Durham DH1 3LE, United Kingdom\\
$^{2}$Department of Physical Sciences, Indian Institute of Science Education and Research Mohali, Sector 81, S.A.S. Nagar, Punjab 140306, India\\
$^{3}$Institute for Computational Cosmology, Department of Physics, Durham University, South Road, Durham DH1 3LE, United Kingdom\\
$^{4}$National Centre for Radio Astrophysics, TIFR, Post Bag No. 3, Ganeshkhind, Pune 411007, India\\
$^{5}$Centre for Space Research, North-West University, Potchefstroom 2520, South Africa\\
$^{6}$RAD@home Astronomy Collaboratory, Kharghar, Navi Mumbai 410210, India
}

\date{Accepted XXX. Received YYY; in original form ZZZ}

\pubyear{\the\year{}}

\begin{document}
\label{firstpage}
\pagerange{\pageref{firstpage}--\pageref{lastpage}}
\maketitle

\begin{abstract}
Tailed radio galaxies are powerful tracers of interactions between active galactic nuclei (AGN) and the intracluster medium (ICM), providing unique insights into cluster dynamics. We present LOw Frequency ARray (LOFAR) 144 MHz and uGMRT 400 MHz observations of the cluster MACS\,J1354.6+7715 (z = 0.3967) to investigate the radio emission associated with its member galaxies and the cluster environment. The dominant tailed radio galaxy in the cluster exhibits a sharply bent tail extending over $\sim300$ kpc, with the spectral index steepening from $\alpha \approx -0.46 \pm 0.21$ near the AGN core to $\alpha \approx -2.43 \pm 0.30$ in the outermost regions. Synchrotron modelling of the tail yields a radiative age of $150\pm10$ Myr, implying a galaxy velocity of $1956 \pm 130$ $\mathrm{km\,s^{-1}}$, which is of order $\sim 0.9$ times the escape velocity. We further found no evidence of relics or halos in our radio images and the X-ray morphology from \textit{Chandra} appears relatively undisturbed, suggesting that the system is in a pre-merging candidate. Our results indicate that the radio galaxy is undergoing its first infall into the cluster, providing an excellent laboratory for studying the impact of the ICM on AGN activity and galaxy evolution, and demonstrating how multi-frequency radio observations of tailed galaxies can uniquely probe both AGN lifecycles and the early stages of cluster assembly.
\end{abstract}




\section{Introduction}
Determining the dynamical state of a galaxy cluster is a prerequisite to understand how the intracluster medium (ICM) regulates the evolution of its member galaxies and their central active galactic nuclei (AGN). In particular, clusters in an immediate pre or post-merger scenario provide key constraints on the impact of cluster merger events as the nascent interaction between substructures can generate shocks, turbulence, and bulk gas motions that influence AGN jet propagation, gas stripping, and star formation activity \citep{feretti_2002,markevitch_2007,vazza_2012}. These processes can heat the ICM, amplify magnetic fields, and accelerate particles to relativistic energies, giving rise to diffuse radio emission in the form of radio halos and radio relics \citep{ferrari_2008,brunetti_2014,gennaro_2023,rajpurohit_2024}.

Radio relics, in particular, are believed to trace merger shocks, visible as arc-like synchrotron structures in radio images and often accompanied by surface brightness discontinuities in X-ray observations \citep{ensslin_1998,weeren_2010,bonafede_2022}. These relics provide direct observational evidence for large-scale shocks propagating through the ICM. However, the absence of detectable shocks does not imply dynamical quiescence: cluster mergers can proceed in stages, and weak or oblique shocks may be difficult to observe, especially in high-redshift or low-surface-brightness systems \citep{Wilber_2018,golovich_2019,Giovannini_2020}. In such cases, other tracers are needed to assess the cluster’s dynamical state and understand its assembly history. \\
\hspace*{1em}Radio galaxies embedded in clusters offer one such probe into the ICM’s physical conditions and the cluster’s dynamical state \citep{feretti_2012a,weeren_2019,ignesti_2022}. A significant fraction of cluster galaxies host AGN \citep{Klesman_2012,Ehlert_2014,Bufanda_2017}, whose relativistic jets interact with the hot ICM. In dense environments, this interaction modulated by the galaxy's motion often leads to the formation of tailed radio galaxies, in which elongated synchrotron emission trails behind the host \citep{fabian_1999,kormendy_2001,merritt_2001}. The morphology of these radio tails, shaped by the motion of the host galaxy through the ICM, offers a direct probe of the galaxy’s kinematics relative to the surrounding gas and can inform us about both the galaxy's trajectory and the state of the ICM itself \citep{pfrommer_2011}. Recent statistical studies have demonstrated strong correlations between the bending angle of radio jets, cluster-centric distance, and galaxy velocity, reinforcing their diagnostic power for probing cluster assembly \citep{jagt_2025}. Spectral studies of such sources further constrain the energy losses of the relativistic electron population and the evolution of AGN activity in the cluster source. 

In this paper, we present the first dedicated multi-frequency radio study of MACS\,J1354.6+7715 (see Table \ref{tab:cluster_properties} for the main cluster properties), detected by the Massive Cluster Survey \citep{ebeling_2001,ebeling_2010}. We have used high-resolution, low-frequency radio observations, focusing on the AGN activity within the cluster and investigating the presence of any diffuse synchrotron or shock-related emission to assess the dynamical state of the cluster and the role of environmental interactions. Despite its inclusion in statistical lensing studies \citep{horesh_2010} and indications of dynamical activity from its X-ray morphology \citep{yuan_2020}, MACS\,J1354.6+7715 has not yet been the subject of detailed multi-wavelength analyses. While \citet{repp_2018} reported signs of star formation in the brightest cluster galaxy (BCG) using HST, no extensive morphological or spectroscopic studies have been
conducted, and the cluster’s non-thermal radio properties remain largely unexplored. The absence of spectroscopic redshift information for the member galaxies also limits our ability to directly assess substructure or galaxy dynamics, underscoring the importance of complementary radio and X-ray diagnostics used in this work.

The paper is organized as follows: in Section~\ref{sec:data&methods} we describe the data and methodology, including details on observations and data reduction techniques; in Section~\ref{sec:results} we present the radio morphology and spectral analysis of the radio tail, including ageing models and age estimation; in Section~\ref{sec:discussion} we interpret the results in a broader physical context using multi-wavelength information; and in Section~\ref{sec:conclusion} we summarize our findings and outline their implications. Throughout this paper, we adopt a flat $\Lambda$CDM cosmology with $H_0=70\,\mathrm{km\,s^{-1}Mpc^{-1}}$, $\Omega_\mathrm{M}=0.3$, and $\Omega_\mathrm{\Lambda}=0.7$. At the redshift of MACS\,J1354.6+7715, 1 arcsecond corresponds to 5.4\,kpc.
\begin{table}
\centering
\caption{Properties of the galaxy cluster MACS\,J1354.6+7715 from \citet{piffaretti_2011}.}
\label{tab:cluster_properties}
\begin{tabular}{lc}
\hline\hline
Property & Value \\
\hline
RA [$^{\circ}$] & 208.6575 \\
Dec [$^{\circ}$] & +77.2597 \\
Redshift $z$ & 0.3967 \\
$M_{500}$ [$\times 10^{14}\ M_{\odot}$] & 5.6900 \\
$R_{500}$ [Mpc] & 1.0960 \\
$L_\mathrm{{X500}}$ [$\times 10^{37}$ W] & 8.7334 \\
\hline
\end{tabular}
\end{table}
\section{Data and Methods} \label{sec:data&methods}
Here we describe the observations, data reduction, and imaging with the LOw Frequency ARray \citep[LOFAR;][]{vanhaarleem_2013} and the upgraded Giant Metrewave Radio Telescope \citep[uGMRT;][]{swarup_1991}.
\subsection{LOFAR}
The HBA (High Band Antenna) observations were conducted as part of the LOFAR Two-Metre Sky Survey \citep[Project ID: LT$14\_004$]{Shimwell_2017,Shimwell_2022}, centered at an observing frequency of 144 MHz with a total bandwidth of 48 MHz (120–168 MHz). The observations were taken on 19 September 2021, starting at 09:41:00.0 UTC and lasting for approximately 8 hours. The data were recorded using 231 sub-bands, each with 16 channels, and a channel width of 12.205 kHz, with 1-second integration times and in both linear polarizations. Following the flagging of bad antennas, a total of 48 stations were used for high-resolution imaging, comprising 23 core stations, 12 remote stations, and 13 international stations. The inclusion of international stations provides the longest projected baseline of $\sim$1432 km, which results in an angular resolution  of around 0.3 arcseconds at 144 MHz. In addition, two primary calibrators, 3C\,196 and 3C\,295, were observed for 10 minutes each, directly before and after the target observation, respectively. 
\subsubsection{Dutch Stations Calibration}
We began by calibrating the shorter baselines in the Dutch array using the \textsc{LINC\footnote{\href{https://git.astron.nl/RD/LINC}{https://git.astron.nl/RD/LINC}}}\textsc calibration pipeline \citep{weeren_2016,williams_2016,gasperin_2019}. The pipeline is divided into two parts: \textsc{LINC Calibrator} and \textsc{LINC Target}. The first part operates only on the data from the two primary calibrators used in this observation, 3C\,295 and 3C\,196. We note that the calibrator stage uses data from all Dutch and international stations to derive direction-independent calibration solutions, including polarization alignment, Faraday rotation, bandpass response, clock offsets, and Total Electron Content (TEC). These solutions are obtained by comparing the observed visibilities to a well-defined sky model of the calibrator source, to correct for slowly varying instrumental and ionospheric effects \citep{gasperin_2019}. Of the two calibrators, 3C\,196 resulted in the solutions with the highest signal-to-noise ratio, and was therefore chosen as the reference calibration solutions for instrumental effects. These are: polarization alignment, bandpass, and clock offset calibration solutions, which were transferred to the target data. The \citet{scaife_2012} (SH12) flux scale was adopted for setting the absolute flux density scale of the observations. 

The calibration solutions derived from 3C196 were applied only to the Dutch station data of the target field via the \textsc{LINC Target} pipeline. This workflow performs an initial direction-independent phase calibration. Further flagging was performed on target data \citep{offringa_2012}, and contamination from bright off-axis sources was mitigated with clipping. Next, the rotation measure (RM) for each station, including international stations, was obtained in the direction of the target field using Global Positioning System (GPS) data via the \textsc{RM Extract} package \citep{mevius_2018}. These corrections, sampled at 15-minute intervals, account for bulk Total Electron Content (TEC) variations and provide the basis for subsequent differential TEC (dTEC) calibration. Finally, a phase-only calibration was performed against a sky model from TGSS (TIFR-GMRT Sky Survey; \citeauthor{intema_2017} \citeyear{intema_2017}).

\begin{figure*}
    \centering
    \includegraphics[width=\linewidth]{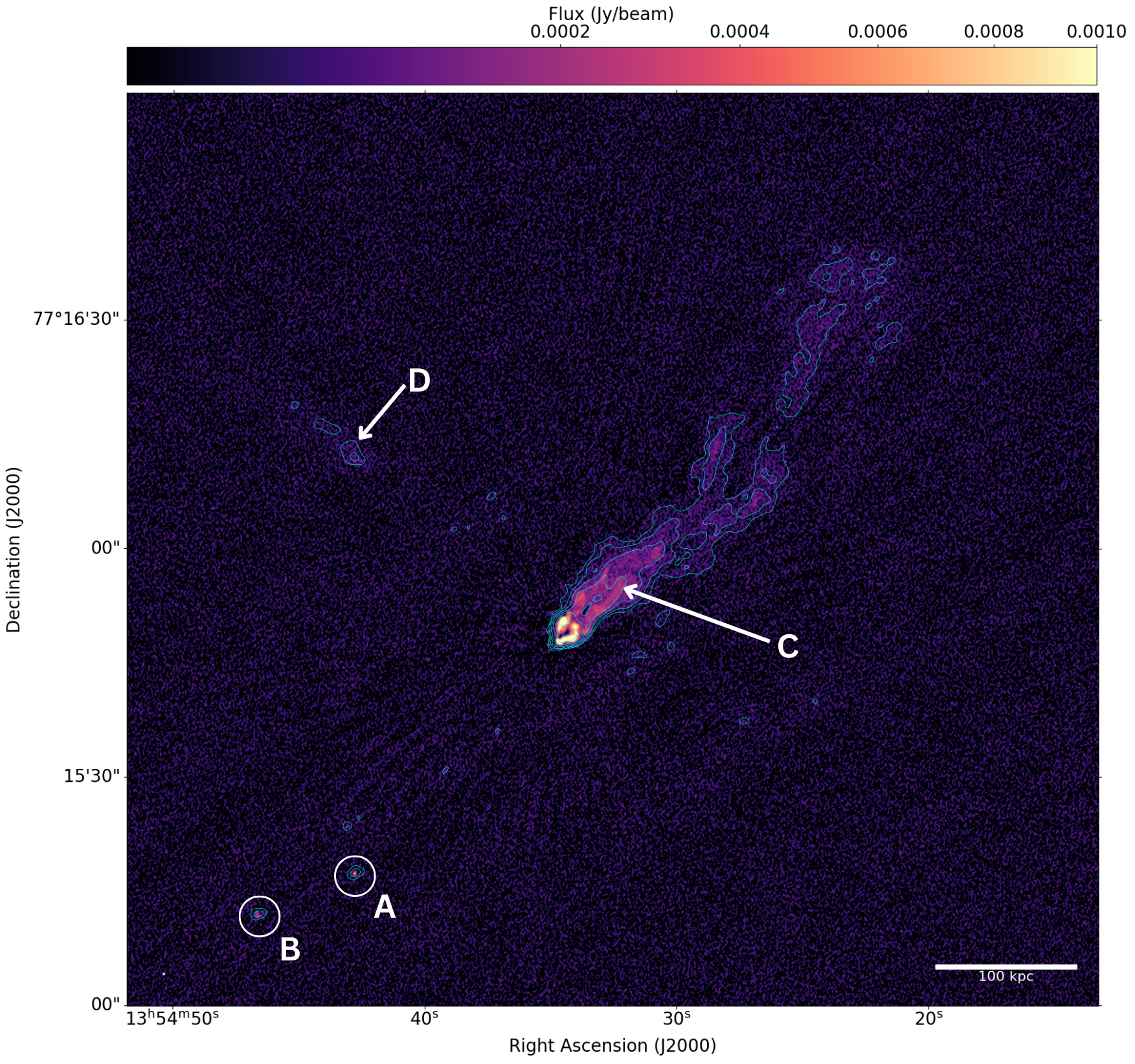} 
    \caption{High-resolution LOFAR HBA image (beam size: $0.30''\times 0.21''$, $\sigma_{\mathrm{rms}} = 0.031\mathrm{mJy\,beam^{-1}}$) of the tailed radio galaxy (C) in the galaxy cluster MACS\,J1354.6+7715, observed at 144\,MHz. A and B denotes two compact radio sources, while D marks an additional radio source detected in the field. The colour scale (square-root scaling, $\gamma = 0.5$) spans from $10^{-7}\,\mathrm{Jy\,beam^{-1}}$ up to $10^{-3}\,\mathrm{Jy\,beam^{-1}}$, revealing both faint diffuse emission and compact features along the tail. Contours from the intermediate-resolution image (beam size: $1.56''\times 1.04''$, $\sigma_{\mathrm{rms}} = 0.099\,\mathrm{mJy\,beam^{-1}}$) are overlaid in cyan, starting at $5\sigma_{\mathrm{rms}}$ and increase by successive factors of $\sqrt{2}$. The white bar in the bottom-right corner corresponds to a physical scale of 100 kpc.}
    \label{Fig:2.1}
\end{figure*}
\subsubsection{International Stations Calibration}
After calibrating the Dutch stations, we proceeded to calibrate the international stations using the LOFAR-VLBI pipeline \citep{Morabito_2022}, which consists of two main steps: delay calibration and self-calibration. The process began by selecting a suitable calibrator from the Long Baseline Calibrator Survey \citep[LBCS;][]{moldon_2015,Jackson_2016}. The pipeline automatically queried the LBCS\footnote{\href{https://lofar-surveys.org/lbcs.html}{https://lofar-surveys.org/lbcs.html}} catalogue within a default radius of $1.5^{\circ}$ from the pointing centre, selected the best suitable in-field calibrator and phase-shifted the data to its direction to begin the calibration process. For our observations, the calibrator identified was the LBCS source L474286 at $\mathrm{R.A} = 208.04296^\circ, \; \mathrm{Dec} = 77.99933^\circ.$ \\
\begin{figure*}
    \centering
    \begin{minipage}{0.49\textwidth}
        \centering
        \includegraphics[width=\linewidth]{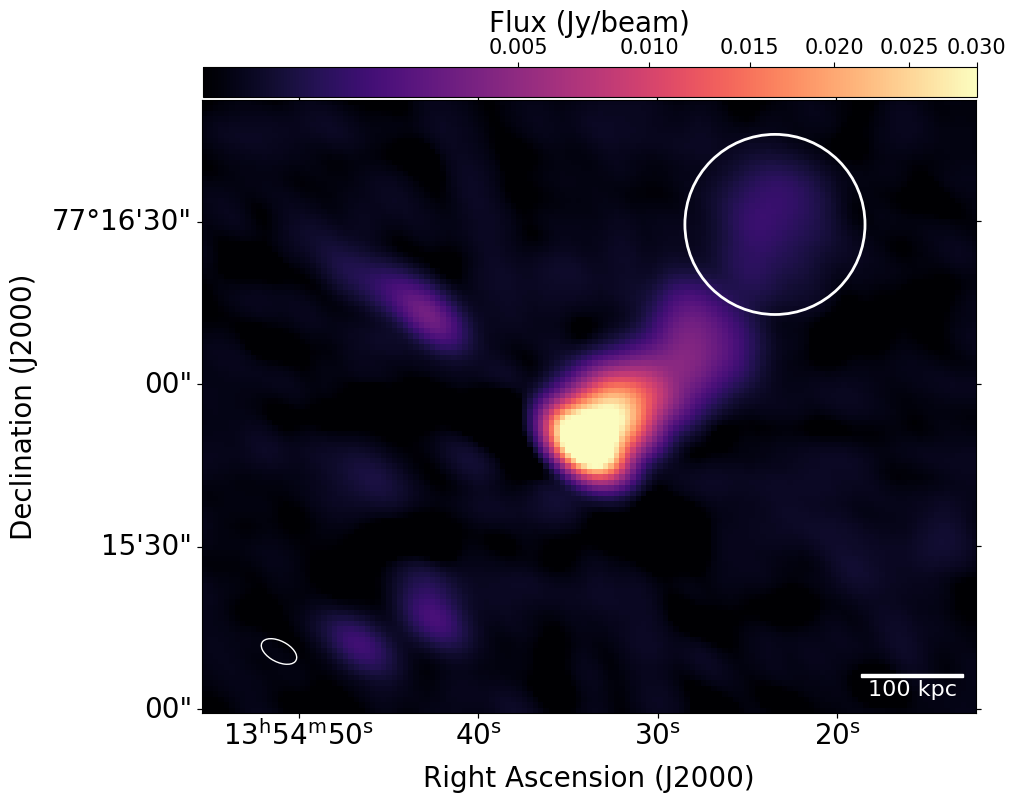}
    \end{minipage}
    \hfill
    \begin{minipage}{0.49\textwidth}
        \centering
        \includegraphics[width=\linewidth]{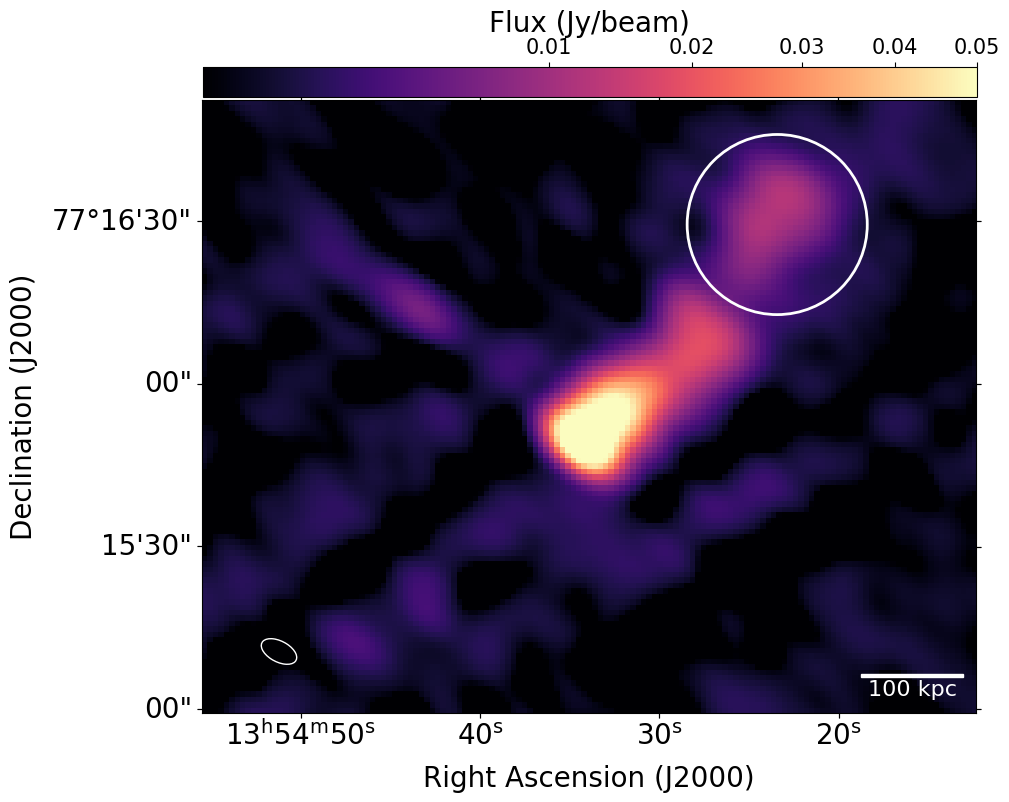}
    \end{minipage}
    \caption{\textbf{Left:} uGMRT Band-3 image (centered at 400\,MHz) of the tailed radio galaxy in MACS\,J1354.6+7715 with a synthesized beam of $9.53'' \times 5.73''$ and an rms noise level of $0.034\,\mathrm{ mJy\,beam^{-1}}$. The colour scale (square-root scaling, $\gamma = 0.5$) spans from $5\times10^{-9}\,\mathrm{Jy\,beam^{-1}}$ to $3\times10^{-2}\,\mathrm{Jy\,beam^{-1}}$.
\textbf{Right:} LOFAR HBA image (centered at 144\,MHz) convolved to match the uGMRT resolution, with an rms noise level of $0.323 \,\mathrm{mJy\,beam^{-1}}$. The colour scale (square-root scaling, $\gamma = 0.5$) spans from $5\times10^{-9}\,\mathrm{Jy\,beam^{-1}}$ to $5\times10^{-2}\,\mathrm{Jy\,beam^{-1}}$.
The white ellipse in the bottom-left corner of each panel indicates the synthesized beam, and the white bar in the bottom-right corner of both panels corresponds to a physical scale of 100\,kpc. 
The white circles (radius $=$ 90\,kpc) mark the regions along the tail used for spectral ageing analysis (see Section \ref{sec:4.2} ).}
    \label{fig:2.2}
\end{figure*}
 \begin{figure*}
    \centering
    \begin{minipage}{0.48\textwidth}
        \centering
        \includegraphics[width=\linewidth]{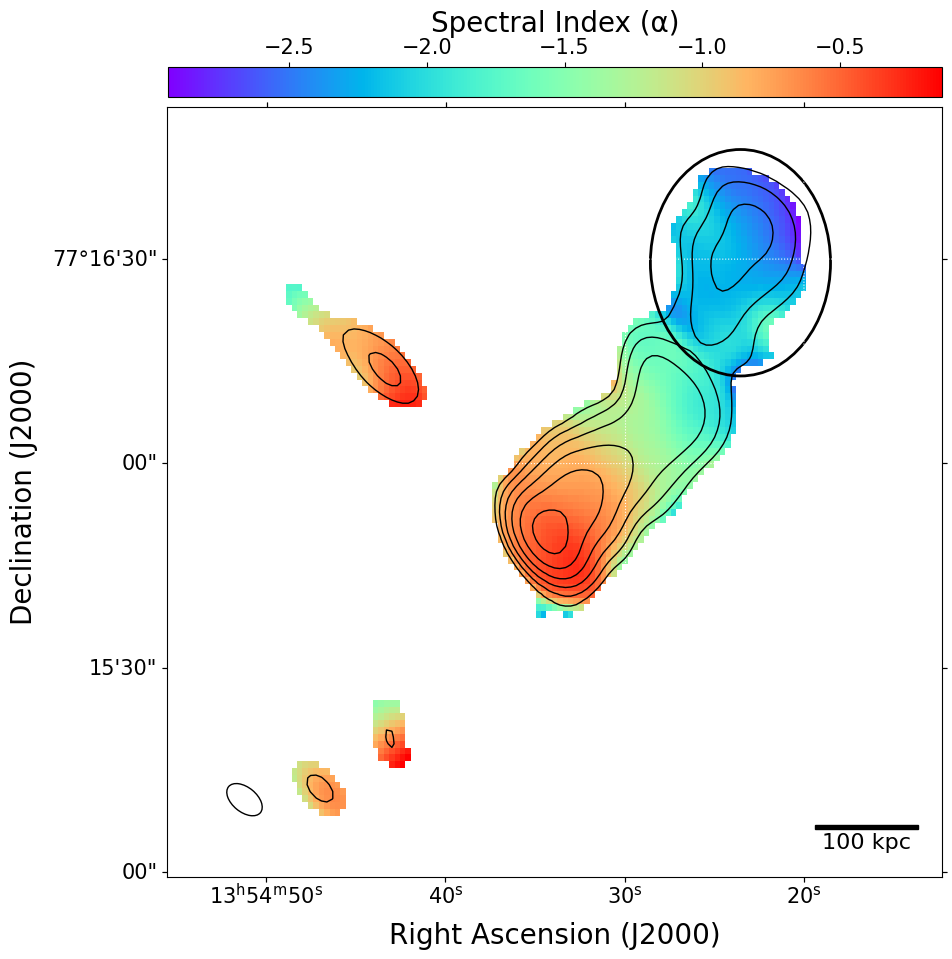}
    \end{minipage}
    \hfill
    \begin{minipage}{0.48\textwidth}
        \centering
        \includegraphics[width=\linewidth]{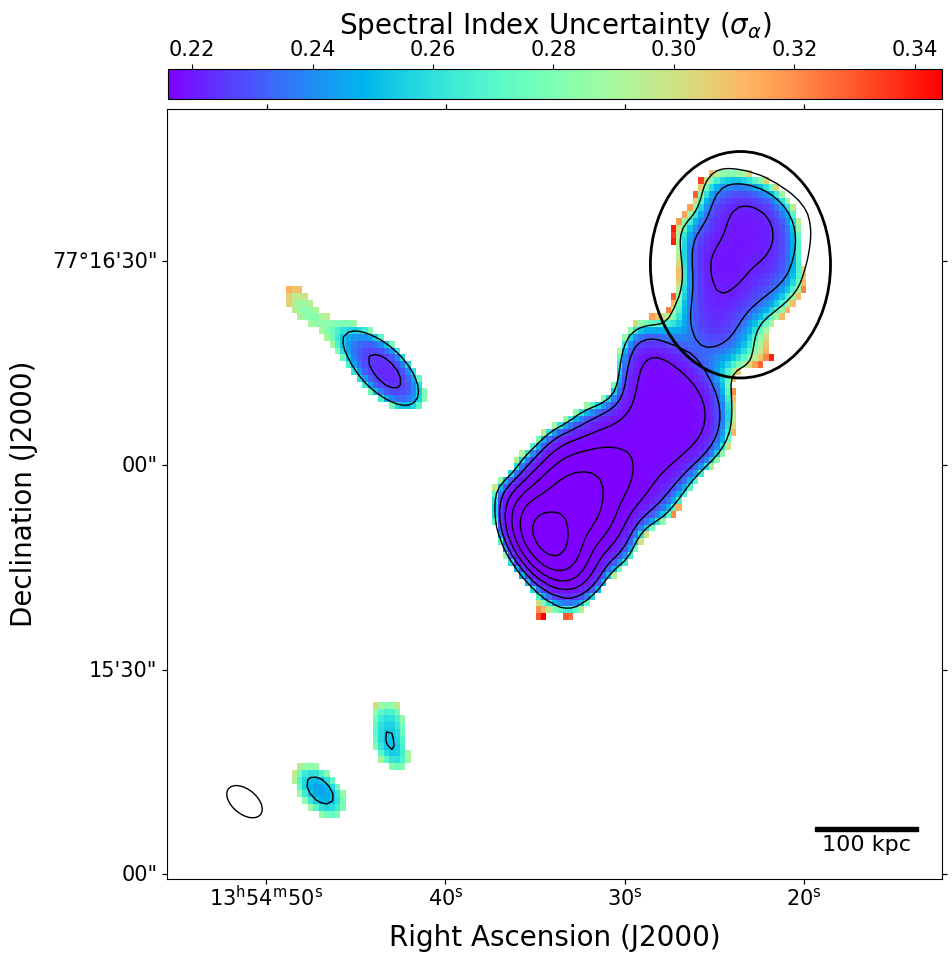}
    \end{minipage}
    \caption{Spectral index map of the radio emission in the cluster region (\textbf{Left)}, with the corresponding uncertainty plot (\textbf{Right}), derived between 144\,MHz and 400\,MHz. The head of the tailed radio galaxy shows a spectral index of $\alpha \approx -0.46 \pm 0.21$, while the spectral index steepens along the tail to values as low as $\alpha \approx -2.43 \pm 0.30$. The median spectral index uncertainty is $\sigma_{\alpha} \approx 0.21$, increasing to values of $\sim 0.35$ toward the edges of the detected emission. The contours show the 144 MHz radio emission as detected by LOFAR (right panel of Fig. \ref{fig:2.2}), matched to the uGMRT resolution. The black contours are drawn at levels starting at $5\sigma_{\mathrm{rms}}$ ($\sigma_{\mathrm{rms}} = 0.323\,\mathrm{mJy\,beam^{-1}}$) and increase by multiplicative factors of $\sqrt{2}$. The synthesized beam is shown as a black ellipse in the bottom-left corner, and the black bar in the bottom-right corner indicates a physical scale of 100 kpc. The black circles (radius = 90 kpc) mark the regions along the tail used for spectral ageing analysis (see Section \ref{sec:4.2} ).}
    \label{Fig:2.3}
\end{figure*}
\hspace*{1em}To reduce data volume while maintaining calibration accuracy, the data were averaged to two channels per subband in frequency and to eight-second intervals in time. This averaging narrows the effective field of view via bandwidth and time smearing \citep{bridle_1999} which helps suppress contributions from nearby sources. A beam correction is then applied to account for the difference between the station beam pattern computed at the original phase center and that in the direction of the selected delay calibrator. During the derivation of the calibration solutions, a minimum baseline length of $50k\lambda$ (where $\lambda$ is the wavelength) was applied to suppress the environment of the in-field calibrator. Following this, the delay-calibration pipeline derived corrections for dispersive delays against a point-source model for the calibrator. The data underwent multiple rounds of self-calibration to iteratively improve the antenna-based phase and gain solutions. In each cycle, a new sky model was generated from imaging and deconvolution, which was then used to refine the phase and gain solutions. Nine rounds of self-calibration were performed, starting with a phase-only calibration and followed by joint phase-and-gain calibration in the subsequent iterations until it converged to the best solution.\\ 
\hspace*{1em}We found minor positional inaccuracies in the coordinates catalogued in LBCS which introduced astrometric offsets. To correct for these, we manually repeated the delay calibration \citep{weeren_2021} using a point-source model constructed from the Very Large Array Sky Survey \citep[VLASS;][]{lacy_2020}, which provided more accurate source positions and significantly improved astrometric precision. In this step, we applied scalarphase calibration while introducing a smoothness constraint, which is regularization term that controls how rapidly the phase solutions are allowed to vary over frequency, helping to suppress solution noise. The level of this constraint was adjusted based on the station type: we used a narrow constraint (smoothness value of 2.0\,MHz) for the international stations, moderate constraints for the remote stations (smoothness 10.0\,MHz), and a wider constraint (smoothness 40.0\,MHz) for the core stations, where ionospheric variations are less pronounced due to shorter baselines. These scalarphase steps were performed over progressively increasing solution intervals (from 32 seconds up to 4 minutes), depending on the baseline length and the station group. 
\begin{figure*}
    \centering
        \centering
        \includegraphics[width=\linewidth]{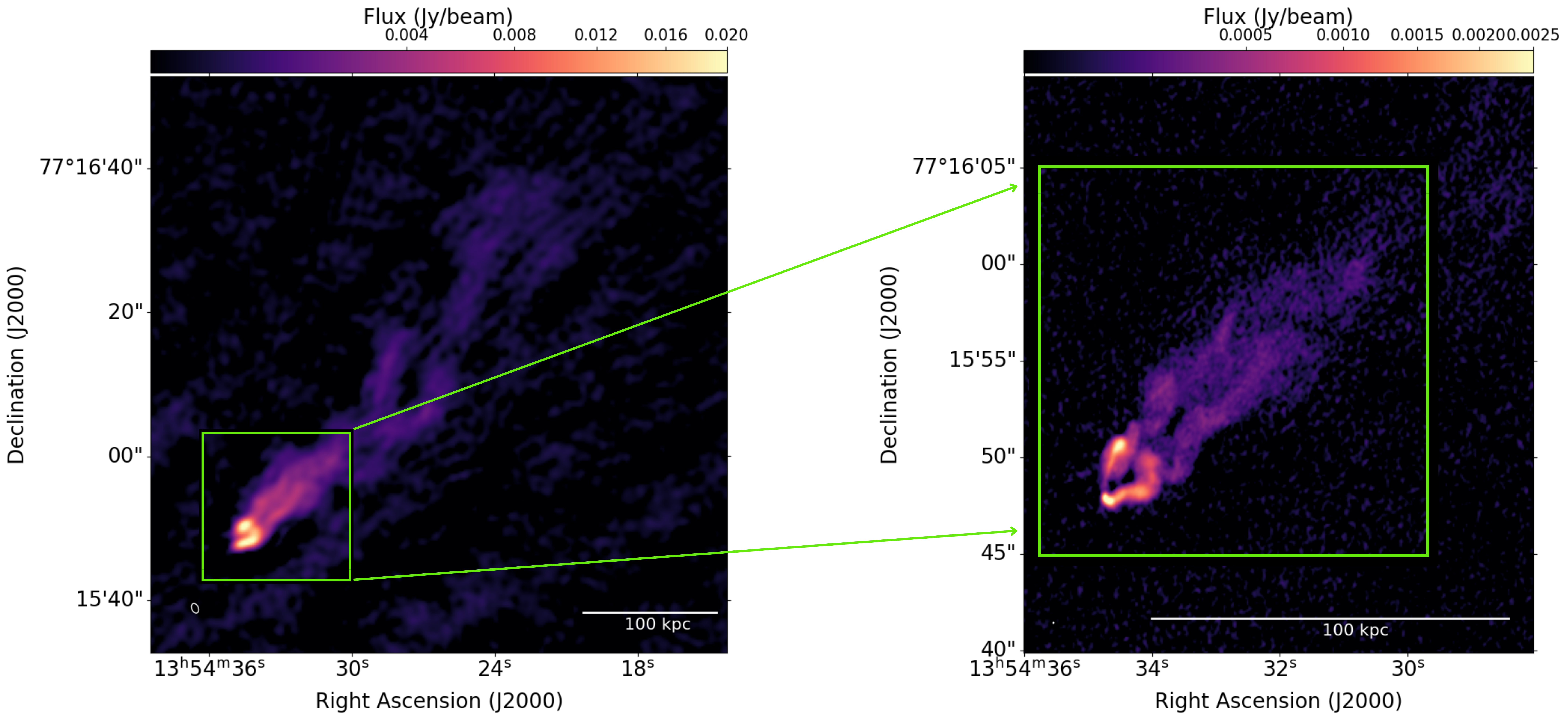}
    \caption{Zoom-in view of the Intermediate-resolution image (\textbf{Left}; beam size $1.56''\times1.04''$, $\sigma_{\mathrm{rms}} = 0.099\,\mathrm{ mJy\,beam^{-1}}$) and High-resolution image (\textbf{Right}; beam size $0.30''\times0.21''$, $\sigma_{\mathrm{rms}} = 0.031\,\mathrm{ mJy\,beam^{-1}}$) highlighting detailed features of the radio jet and tail. The colour scale (square-root scaling, $\gamma = 0.5$) spans from $3\times10^{-5}\,\mathrm{Jy\,beam^{-1}}$ to $2.5\times10^{-2}\,\mathrm{Jy\,beam^{-1}}$ in the high-resolution image and from $5\times10^{-5}\,\mathrm{Jy\,beam^{-1}}$ to $2\times10^{-2}\,\mathrm{Jy\,beam^{-1}}$ in the intermediate-resolution image. In both panels, the white ellipse in the bottom-left corner indicates the synthesized beam, and the white bar in the bottom-right corner corresponds to a physical scale of 100 kpc.}
    \label{fig:3.1}
\end{figure*}

Before self-calibration on the target, the delay-calibrator solutions were applied to the full international data, along with the Dutch phase corrections obtained from the earlier calibration step. The data were then phase-shifted to the target source. This was done using the \textsc{Split-Directions} pipeline. The pipeline solves for dTEC solutions, since the ionosphere conditions are generally different between the target and the delay calibrator. This was followed by several rounds of both phase-only and phase and amplitude self-calibration. At each step, the data were re-imaged using \textsc{WSClean} \citep{offringa_2014} with Briggs weighting (robust = $-1.5$). The calibration process was repeated until the signal-to-noise ratio in the resulting images no longer showed a substantial improvement. However, the output images from \textsc{Split-Directions} exhibited artefacts indicative of residual calibration errors. To address this, we applied a highly conservative manual self-calibration, solving for ionospheric corrections during self-calibration. Lastly, to fully deconvolve and model the extended emission in the target, we used multiscale cleaning.

To better recover the extended, low-surface-brightness emission, we created an intermediate-resolution image of the target field. For this, we used the calibrated data obtained from the high-resolution self-calibration step and applied a gaussian uv-taper of 1.2 arcseconds during imaging with \textsc{WSClean}. This approach allowed us to enhance sensitivity to diffuse structures in the AGN tail while maintaining sufficient resolution for morphological analysis. Figure \ref{Fig:2.1} shows the final calibrated high-resolution image with the contours of the intermediate-resolution image overlaid. The uncertainty on the absolute flux scale is $20\%$ for the LOFAR data \citep{Morabito_2022}.
\subsection{uGMRT}
We also used archival uGMRT Band-3 (400\,MHz) data (PI: Sravani Vadi, $37\_128$) for our spectral analysis. The cluster was observed in GMRT Wideband Backend (GWB) mode for the duration of $\sim$6 hours, from 16 December 2019 21:54:05 TAI to 17 December 2019 03:11:39 TAI. The data were recorded in 4096 spectral channels, with a 2.6-second integration time across a bandwidth of 200\,MHz. The primary and secondary calibrators observed were 3C286 and 1459+716, respectively.  
\subsubsection{Data Calibration}
The uGMRT data were processed using the Python-based continuum imaging pipeline CAPTURE \citep{kale_2020}, which utilizes tasks from Common Astronomy Software Applications \citep[CASA;][]{mcmullin_2007}. The data reduction included RFI mitigation, calibration using standard calibrators, and subsequent self-calibration. 3C286 was used for flux density, delay, and bandpass calibration, while 1459+716 served as the phase calibrator. The \citet{perley_2017} (PB17) flux scale was adopted. For consistency across datasets, the LOFAR image was later corrected to match this flux scale (see Section \ref{sec:2.3}) . To deal with RFI, the CASA task \textsc{flagdata} was applied in \textsc{tfcrop} mode before calibration, and a combination of \textsc{rflag} and \textsc{tfcrop} was used between self-calibration iterations. Baseline-based cutoffs in \textsc{rflag} helped preserve short baselines to retain extended emission.
\begin{figure*}
    \centering    
    \begin{minipage}{0.49\textwidth}
        \centering
        \includegraphics[width=0.8\linewidth]{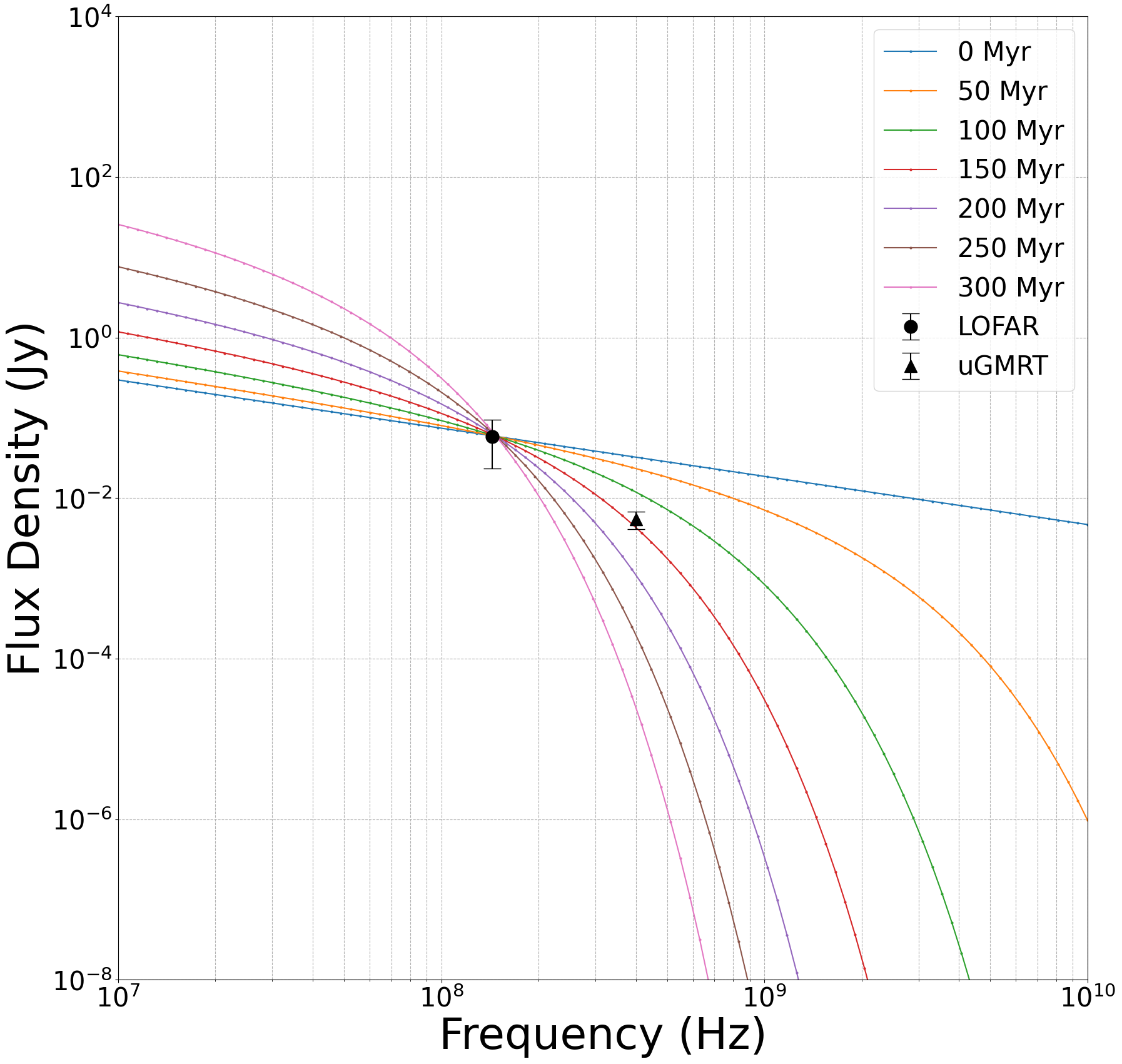}
    \end{minipage}
    \hfill
    \begin{minipage}{0.49\textwidth}
        \centering
        \includegraphics[width=0.8\linewidth]{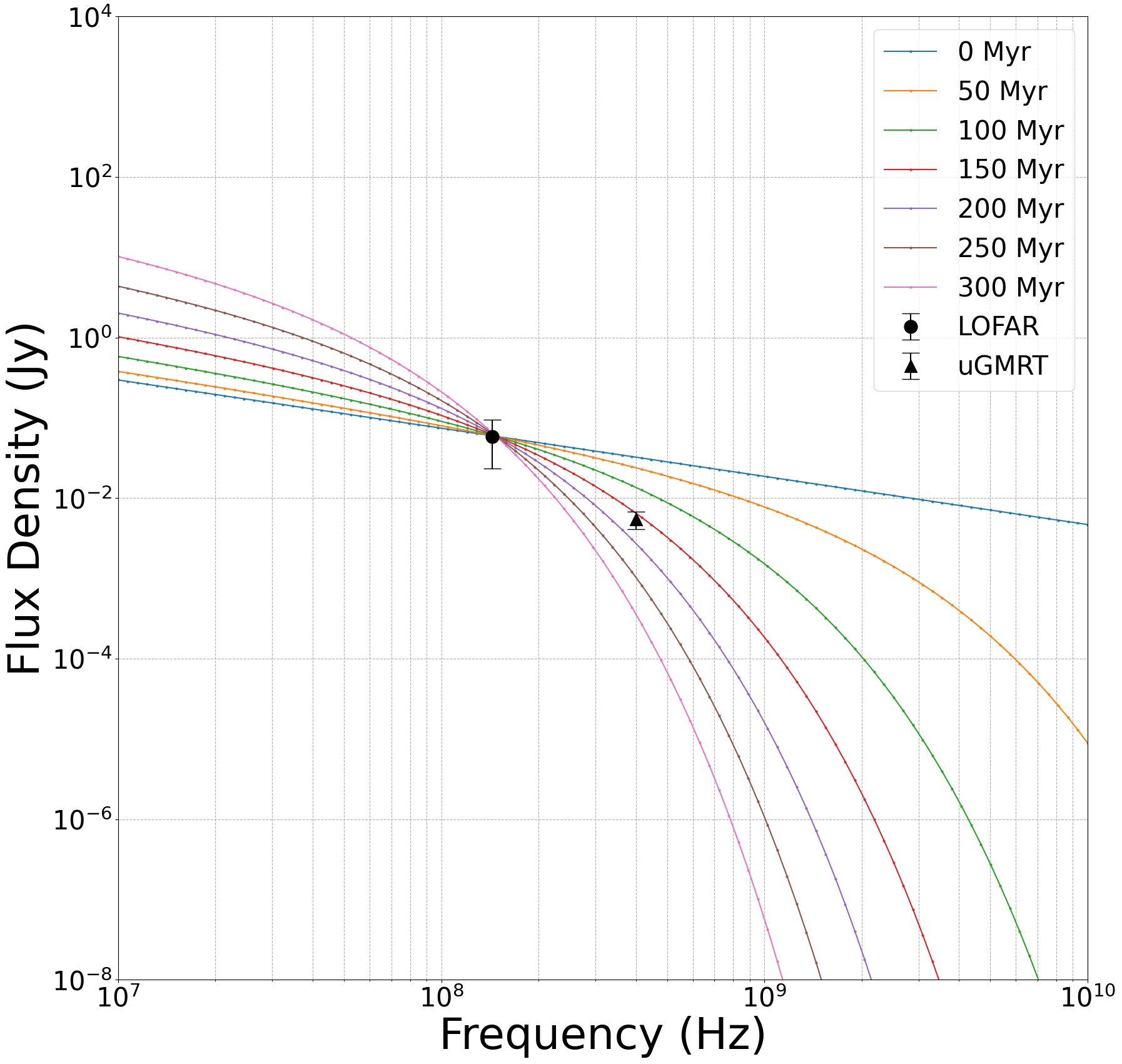}
    \end{minipage}
    
    \caption{ Spectral ageing fits using two different models: shows the Jaffe–Perola (JP) model (\textbf{left}), and the Tribble model (\textbf{right}). The curves represent model spectra for different spectral ages, with the observed LOFAR and uGMRT flux density measurements (over the region mentioned in Fig. \ref{fig:2.2}) overlaid for comparison. The magnetic field assumed in both the fittings is $B=\frac{B_{\mathrm{cmb}}}{\sqrt{3}}$.}
    \label{fig:3.3}
\end{figure*}

Following the initial calibration, the data were split and frequency-averaged with bandwidth smearing in mind and divided into $\sim$6\,MHz subbands to improve convergence during self-calibration.  Eight cycles of self-calibration were carried out using \textsc{tclean}, with Briggs weighting (robust = 0) and $\mathrm{nterms} = 2$. The process included four iterations of phase-only calibration followed by four iterations of joint phase-and-amplitude calibration. In each stage, progressively shorter solution intervals were used, decreasing from 8 minutes to 1 minute during the phase-only stage and from 4 minutes to 1 minute during the phase-and-amplitude stage, to refine the gain solutions. After each round, the residual RFI was flagged, and the model was updated. A primary beam correction was applied to the final image to ensure accurate flux density measurements. Figure \ref{fig:2.2} presents the final calibrated uGMRT image and the LOFAR image smoothed to the same resolution for the later spectral index analysis. We assumed an uncertainty on the absolute flux scale for the uGMRT data of $10\%$ \citep{intema_2017}.
\subsection{Spectral Index Map}\label{sec:2.3}
We constructed the spectral index map by combining LOFAR observations at 144\,MHz with uGMRT data at 400\,MHz, matched in minimum uv-distance and beam size. Differences in the maximum uv-coverage were mitigated by applying a Gaussian uv-taper during imaging. To place both datasets on the same flux scale, the LOFAR image, originally calibrated on the SH12 scale, was rescaled to match the PB17 scale used for uGMRT. We adopted the PB17 scale rather than SH12, as it provides a more recent and widely used standard that extends over a broader frequency range (50\,MHz to 50\,GHz), thereby allowing a uniform calibration between the two datasets. A flux scaling factor of 0.9696 was applied to the LOFAR image to correct for the slight differences between the two scales at the observing frequency. Pixels in both images were masked using a 5$\sigma$ signal-to-noise threshold to exclude low-S/N regions. The spectral index $\alpha$, defined such that $S\propto\nu^\alpha$, where $S$ is the flux density and $\nu$ is the frequency, was then calculated using the standard logarithmic ratio of flux densities across the two frequencies (see Section \ref{sec:3.2} for details; Fig.~\ref{Fig:2.3}).
\section{Results}\label{sec:results}
In this section, we present the results of our multi-resolution radio imaging and spectral analysis of the galaxy cluster MACS\,J1354.6+7715, with a focus on the embedded radio tail galaxy.
\subsection{Radio morphology}\label{sec:3.1}
The LOFAR high-resolution image ($0.3''$, see Fig. \ref{Fig:2.1}) reveals a rich radio environment in the central region of the cluster. A compact radio source (marked as A) coincides with the BCG near the cluster center. Approximately 1 arcminute to the east, another compact radio source (marked as B) is visible, showing no clear extended emission. Toward the northwest lies the most striking feature in the field, a narrow-angle tailed \citep[NAT;][]{miley_1980,owen_1985} radio galaxy (marked as C).  It exhibits a highly collimated and sharply bent tail with a projected length of approximately 300 kpc. The core of the target galaxy appears compact and bright, typical of an AGN, while the emission becomes increasingly diffuse along the tail, as seen more clearly in zoom-in images shown in Fig. \ref{fig:3.1}.

\begin{figure*}
    \centering
    \includegraphics[width=\textwidth]{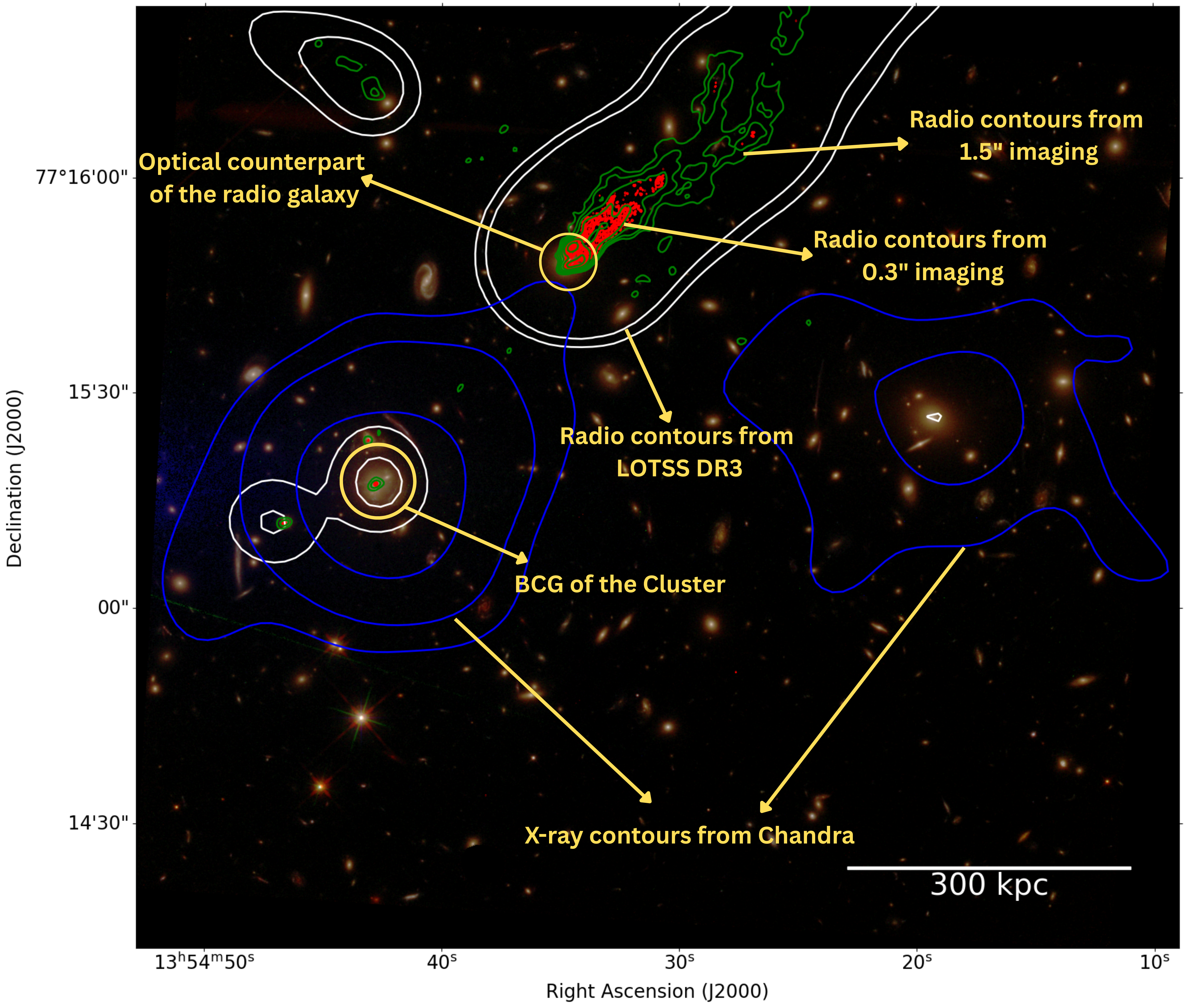}
    \caption{Composite image of the cluster combining optical, radio, and X-ray data. The background Hubble Space Telescope (HST) image shows the galaxy distribution, including the BCG and the optical counterpart of the tailed radio galaxy. Overlaid contours show the 144\,MHz radio emission detected by LOFAR at three angular resolutions: high-resolution ($0.30''\times0.21''$, $\sigma_{\mathrm{rms}} = 0.031\, \mathrm{mJy\,beam^{-1}}$) red contours highlight the fine-scale jet structure, intermediate-resolution ($1.56''\times1.04''$, $\sigma_{\mathrm{rms}} = 0.099\,\mathrm{mJy\,beam^{-1}}$) green contours trace the more diffuse tail, and the LoTSS DR3 image ($6.0''\times6.0''$, $\sigma_{\mathrm{rms}} = 0.075\,\mathrm{mJy\,beam^{-1}}$) white contours capture the large-scale structure of the source. The high-resolution and the intermediate resolution contours are drawn at the same levels as in Fig.~\ref{Fig:2.1}. The two contours level for LoTSS DR3 data correspond to $\sqrt{2}\times5\sigma_{\mathrm{rms}}$ and $2\times5\sigma_{\mathrm{rms}}$. The blue X-ray contours ($\sigma=0.035\,\mathrm{counts\,s^{-1}\,pixel^{-1}}$) extracted from \textit{Chandra} observations, mark the hot ICM, with the contours level corresponding to $17\sigma$, $23\sigma$ and $29\sigma$ (refer to \citet{yuan_2020} for more information regarding \textit{Chandra} image). The white bar in the bottom-right corner corresponds to a physical scale of 300 kpc.}
    \label{Fig:4.1}
\end{figure*}

The tail exhibits multiple bends and wiggles, which may result from a combination of external and internal factors. Externally, fluctuations in ICM density, variable ram pressure as the host galaxy moves through the cluster \citep{bliton_1998,vacca_2022}, and bulk motions such as sloshing or turbulence \citep{morsony_2013} can all affect the jet's path. Internally, variations in the jet power or changes in its orientation may also contribute to the observed structure but the large-scale morphology strongly suggests an external interaction between the AGN jets and the ambient ICM. Further, the tail gradually broadens and fades in brightness, hinting at energy losses due to synchrotron ageing of the relativistic electrons \citep{murgia_1999,parma_1999}, a point we explore in detail through spectral analysis in the following sections. Finally, a faint compact radio source (marked as D) is also detected to the northeast of the radio tail, which is coincident with an optical galaxy in the composite image (refer to Fig.\,\ref{Fig:4.1} under Section \ref{sec:4.1} for more details), showing weak extended emission but lacking any pronounced tailed morphology as seen in source C, and is therefore not discussed further.

\subsection{Spectral modelling and radiative age estimation}\label{sec:3.2}
A spatially resolved spectral index map reveals how the radio spectrum evolves along the tail through synchrotron ageing. As relativistic electrons age due to synchrotron and inverse Compton losses, their energy distribution steepens, resulting in a progressively steeper radio spectrum at higher frequencies. This steepening arises because the radiative loss rate scales as $\left(\frac{dE}{dt}\right)_{rad}\propto E^2B^2$, leading higher-energy electrons, which emit at higher frequencies, to lose energy more rapidly than their lower-energy counterparts, causing the radio spectrum to bend or ``break'' more sharply at high frequencies.   
\begin{figure*}
    \centering
    \includegraphics[width=0.9\textwidth]{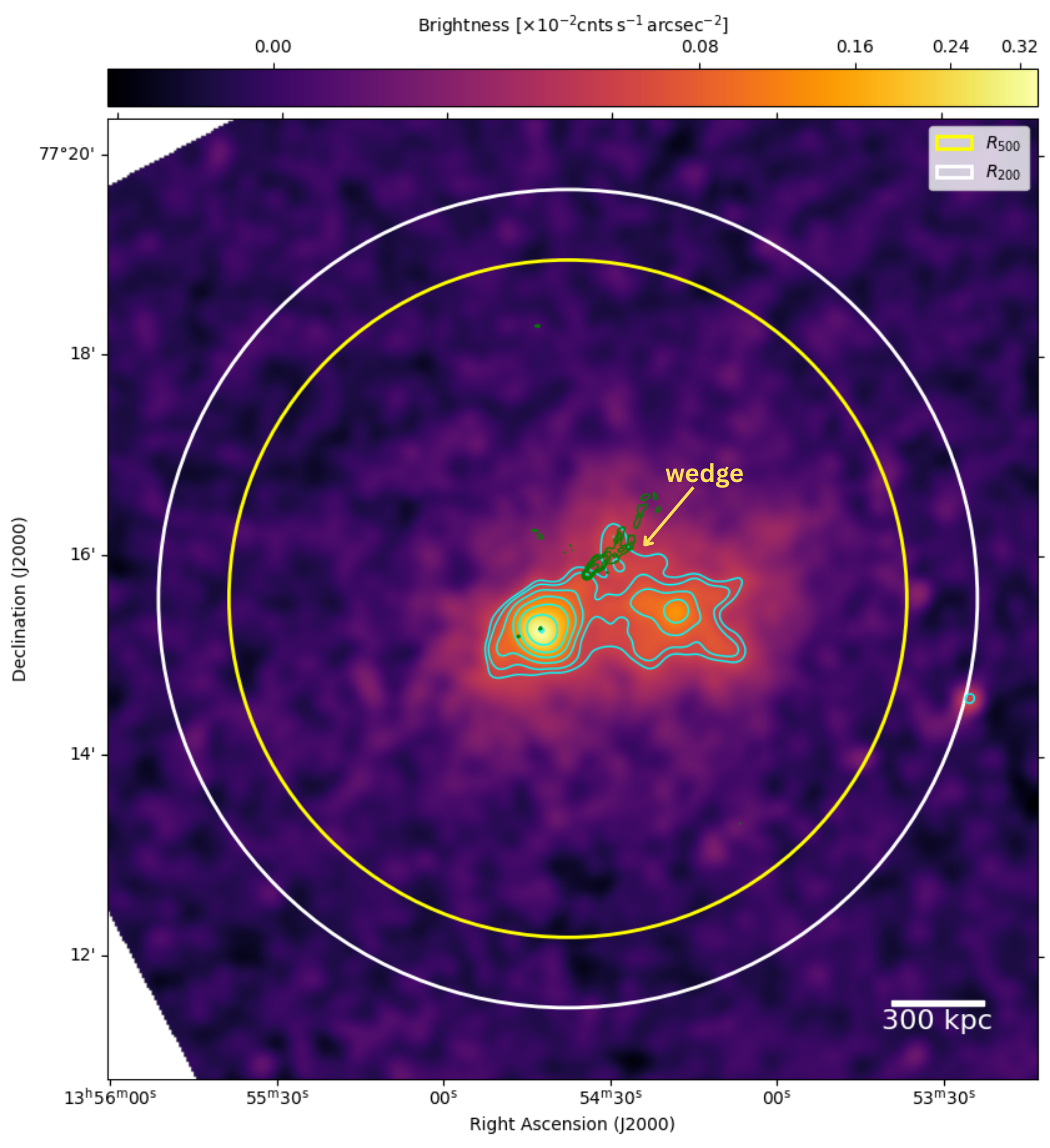}
    \caption{The $3\sigma$ smoothed \textit{Chandra} image from \citep{yuan_2020}. Cyan contours show the X-ray distribution of the central region. The contours are drawn at levels of 5, 6, 8, 11, 15, 23, and 34 times the local average background value of $0.01\,\mathrm{counts\,s^{-1}\,arcsec^{-2}}$, using the ASINH scale. The radio contours, shown in green, are drawn at the same levels as in Fig.\,\ref{Fig:2.1}. The white bar in the bottom-right corner corresponds to a physical scale of 300 kpc.
}
    \label{Fig:4.4}
\end{figure*}
We generated the spectral index map using matched-resolution LOFAR (144\,MHz) and uGMRT (400\,MHz) images. The spectral index was then computed pixel by pixel using:
\begin{equation}
    \alpha=\frac{\log (S_{\nu_{1}}/S_{\nu_{2}})}{\log (\nu_1/\nu_2)} \pm \frac{1}{\log (\nu_1/\nu_2)}\sqrt{\left(\frac{\sigma_{S,\nu_1}}{S_{\nu_1}}\right)^2+ \left(\frac{\sigma_{S,\nu_2}}{S_{\nu_2}}\right)^2 }
\end{equation}
where $\sigma_S$ represents the uncertainty in the flux density at the two frequencies, and $\nu_1$ and $\nu_2$ correspond to 144\,MHz and 400\,MHz, respectively, the frequencies used to generate the spectral index map. The flux density uncertainty is calculated using
\begin{equation}
    \sigma_S = \sqrt{(f \cdot S)^2 + \sigma_{\mathrm{rms}}^2},
\end{equation}
where $f$ denotes the uncertainty in the flux density scale, and $\sigma_{\mathrm{rms}}$ is the local image noise. The resulting spectral index map (Fig.~\ref{Fig:2.3}) clearly reveals spectral steepening along the tail. Near the radio core, we find a relatively flat spectral index of $\alpha \approx~-0.46\pm0.21$, consistent with recent or ongoing particle acceleration. Moving downstream, the tail shows progressively steeper indices, reaching $\alpha \approx -2.43\pm0.30$ in its outermost regions. This gradient is consistent with ageing plasma being advected away from the central engine, as expected from synchrotron energy losses.

To estimate the radiative age of the tail, we have used the \textsc{BRATS}\footnote{\href{http://www.askanastronomer.co.uk/brats/}{http://www.askanastronomer.co.uk/brats/}}
  \citep[Broadband Radio Analysis ToolS;][]{harwood_2013,harwood_2015} to fit the observed flux densities at the two available frequencies with theoretical synchrotron ageing models. While we initially attempted pixel-wise spectral fitting, the limited number of frequencies made the fits underconstrained. Instead, assuming that the spectral age increases with distance along the tail, we measured the integrated flux densities within circular regions of radius of $\mathrm{90\,kpc}$ (corresponding to $16.7''$; see Fig.~\ref{fig:2.2}). This region size was chosen to encompass the full transverse extent of the radio tail at both frequencies while maintaining sufficient signal-to-noise across the  spectral index maps, and was used to constrain the model particularly  at the tail end, where spectral steepening is most pronounced and the oldest electron populations are expected. 
  
  To further reduce the number of free parameters, we assumed a constant magnetic field strength of $B=~B_{cmb}/\sqrt{3}$, where $B_{cmb}=3.25(1+z)^2$~$\mu \mathrm{G}$ is the equivalent magnetic field strength of the cosmic microwave background \citep{miley_1980}. This choice is motivated by the fact that in the outer regions of galaxy clusters (away from the optical BCG), where the radio tail is located, the cluster’s own magnetic field is expected to be weak. As a result, the dominant magnetic field contribution comes from the cosmic microwave background (CMB), which sets a natural baseline for radiative losses. We note that adopting $B=B_{cmb}/\sqrt{3}$ corresponds to the magnetic field strength that maximises the radiative lifetime of synchrotron-emitting electrons for a given break frequency. If the true magnetic field were higher, which can be due to amplification by jet-driven turbulence or compression within the ICM, the inferred radiative ages would decrease accordingly. Therefore, our age estimates should be regarded as an upper limit on the true synchrotron age of the plasma.
  
   Although the spectral index measured near the core is relatively flat ($\alpha\approx0.46\pm0.21$), this value does not necessarily represent the true injection index of the relativistic electron population. At the core region, processes such as synchrotron self-absorption and low-frequency free-free absorption come into play, which together with projection effects can flatten the observed spectrum. Also, with only two observing frequencies available, allowing the injection index to vary freely lead to underconstrained fits. We therefore adopted a fixed injection spectral index of $\alpha_\mathrm{{inj}}=-0.6$, which is commonly observed in FR I and head–tail radio galaxies and widely adopted in spectral ageing studies \citep{Murgia_2011,harwood_2013,harwood_2015,turner_2018}.

\begin{figure*}
    \centering    
    \begin{minipage}{0.49\textwidth}
        \centering
        \includegraphics[width=\linewidth]{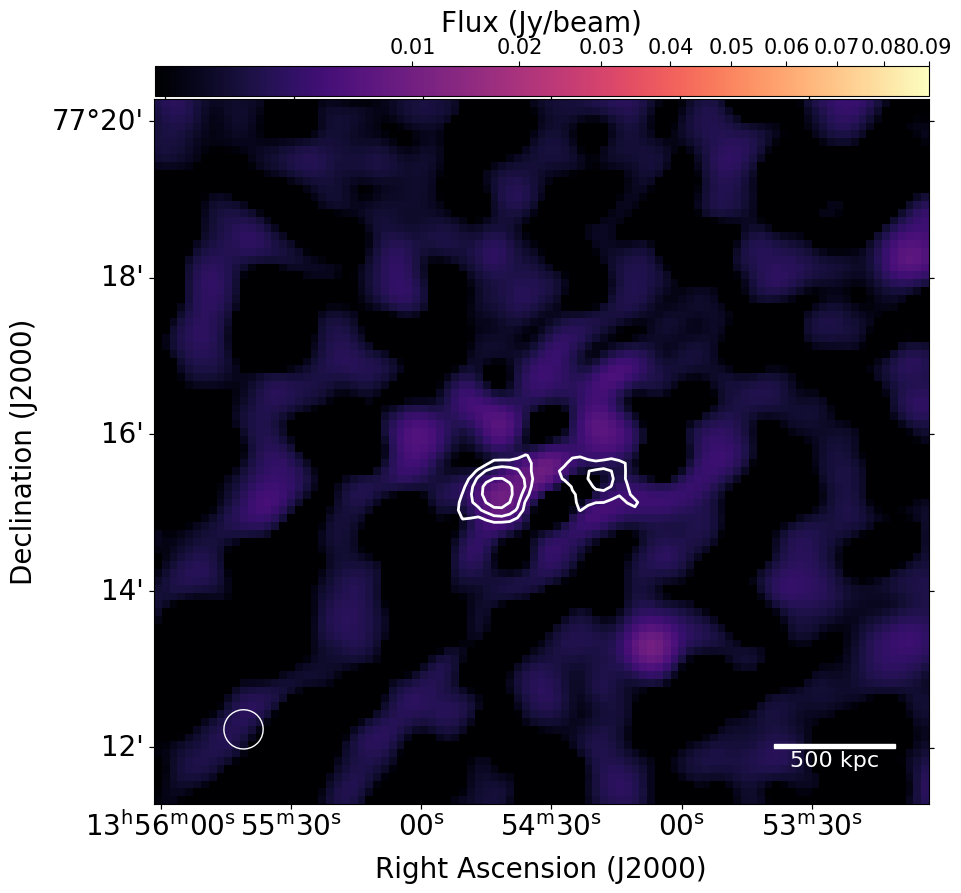}
    \end{minipage}
    \hfill
    \begin{minipage}{0.49\textwidth}
        \centering
        \includegraphics[width=\linewidth]{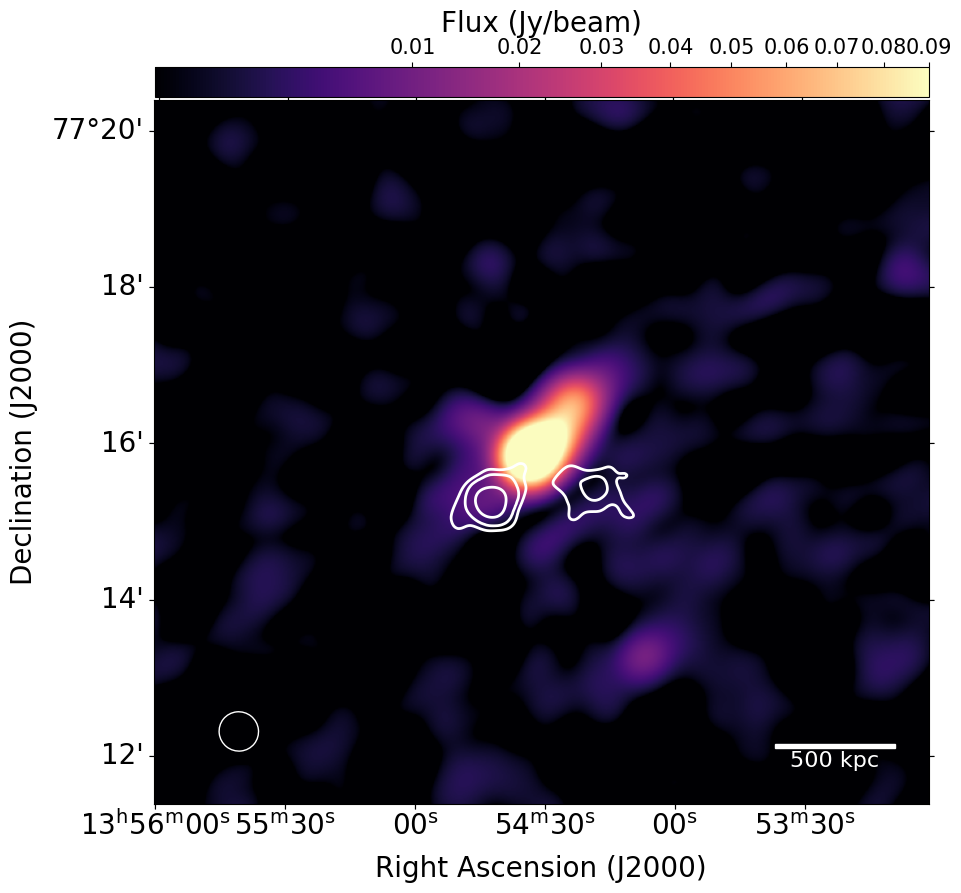}
    \end{minipage}
    
    \caption{\textbf{Left:} Residual LOFAR image of MACS\,J1354.6+7715 (beam size $30.36'' \times 30.08''$, $\sigma_{\mathrm{rms}} = 0.98\,\mathrm{mJy\,beam^{-1}}$) after subtraction of the main radio tail and the smaller eastern tail component, with X-ray contours overlaid as described in Fig.~\ref{Fig:4.1}. \textbf{Right:} LOFAR image including all emission ($\sigma_{\mathrm{rms}} = 1.05\,\mathrm{mJy\,beam^{-1}}$), shown with the same X-ray contours. Both maps are on the same resolution and color scale, which span from $5\times10^{-6}\,\mathrm{Jy\,beam^{-1}}$ to $90\times10^{-3}\,\mathrm{Jy\,beam^{-1}}$. The synthesized beam as a circle and a 500 kpc scale bar are shown in the bottom-left and bottom-right corner of both panels.}
    \label{fig:4.2}
\end{figure*}

Under these assumptions, we fit both the Jaffe–Perola \citep[JP;][]{jaffe_1973} and Tribble \citep{tribble_1993,hardcastle_2013} models to the LOFAR and uGMRT flux densities. The JP model represents an idealized ageing scenario with constant magnetic field and continuous pitch-angle scattering, while the Tribble model accounts for a more realistic, turbulent magnetic field by adopting a Maxwellian distribution of field strengths. These models account for synchrotron and inverse-Compton losses only, with the latter naturally included through the equivalent magnetic field strength of the cosmic microwave background. Other energy-loss processes, such as adiabatic expansion, are not explicitly included in the JP or Tribble models as it would require to model the uncertain expansion of the individual tail region which is out of the scope of this paper. As shown in Fig.\,\ref{fig:3.3}, both models yield consistent radiative ages of $150 \pm 10\,\mathrm{Myr}$, within the uncertainties. The agreement suggests that our spectral age estimate is robust with respect to different assumptions about the magnetic field structure.
\section{Discussion}\label{sec:discussion}
In the previous sections, we presented deep multi-resolution radio imaging of the merging galaxy cluster MACS\,J1354.6+7715. Our observations revealed a striking NAT radio galaxy with a curved and extended tail structure. Through spectral index mapping and spectral ageing analysis, we examined the evolution of the relativistic electron population along the tail and estimated the corresponding radiative age of the plasma. These analyses provide the foundation for understanding how the radio emission traces the dynamical and environmental conditions within the cluster.\\
\hspace*{1em}A central question arising from our observations is why the tailed radio galaxy exhibits such prominent radio emission compared to the rest of the cluster population. At 144\,MHz, the NAT has an integrated flux density of $432.20\pm86.44$ mJy, nearly $\sim$318 times brighter than the BCG, which shows only $1.36\pm0.27$ mJy despite residing at the center of the X-ray bright ICM. This pronounced contrast may reflect substantial differences in AGN activity or gas content between the two galaxies, potentially pointing to a more efficient fueling mechanism in the target galaxy host. Given its location outside the dense cluster core (projected distance of $\sim 219$\,kpc, corresponding to $\mathrm{\sim0.16\,R_{200}}$, from the BCG), it is plausible that this galaxy has retained a larger gas reservoir, which may continue to feed the central black hole. In contrast, the BCG, despite its location at the cluster center, may be experiencing a more quiescent AGN phase. This is consistent with the broader class of tailed radio galaxies (narrow-angle tail/head–tail systems) which are typically associated with non-dominant, infalling cluster galaxies rather than centrally dominated galaxies (cDs). Their morphology, characterised by a bright head at the galaxy nucleus and bent synchrotron tails shaped by ram pressure from motion through the ICM, often places them at large clustercentric radii and high relative velocities \citep[e.g.][]{miley_1975,miley_1980,giacintucci_2009,vos_2021}. Thus, it is not unprecedented for such sources to dominate the radio output of a cluster: classical examples include NGC\,1265 and 3C\,129 \citep[e.g.][]{miley_1975,gendronmarsolais_2020,bhukta_2022}. 

While these findings already underscore the unusual nature of the tailed radio galaxies, a fuller understanding requires placing them in the broader context of their host cluster environment. To that end, we now turn to a multi-wavelength analysis of MACS\,J1354.6+7715 to investigate the dynamical state of the cluster and the role of environmental effects. We then explore how the morphology and location of the radio tail may relate to the galaxy’s motion through the ICM. Finally, we assess how the radiative age derived from spectral modelling compares with expectations from dynamical timescales, and consider the implications for the applicability of standard spectral ageing models.
\subsection{Cluster Environment}\label{sec:4.1}
By combining radio, X-ray, and optical data, we examined the interplay between the tailed radio galaxy, the thermal ICM, and the large-scale gravitational potential of the host system. Figure~\ref{Fig:4.1} presents a composite image of the cluster source.
\subsubsection{X-ray morphology}
For the X-ray data, we used the processed and exposure-corrected images provided by \citet{yuan_2020}, who analysed the dynamical state of galaxy clusters using archival \textit{Chandra} observations. The data were taken directly from their published sample, which included point-source subtraction to isolate diffuse cluster emission followed by image smoothing, and were used without any further reprocessing (see Fig.\,\ref{Fig:4.4}). 

This X-ray emission exhibits  a bimodal surface-brightness distribution with peaks of unequal magnitude suggesting unequal mass merger separated by $\sim450\,\mathrm{kpc}$. The brighter peak coincides with the position of the BCG, suggesting it traces the main cluster core, while the fainter peak lies towards the west, likely associated with a BCG of a secondary, infalling subcluster. Although differences in peak brightness can arise from several factors, such as projection effects, gas density variations, or temperature structure, the presence of two distinct X-ray maxima generally indicates an ongoing interaction between them. Both systems seem rather symmetrical and relaxed with obvious extended emission beyond their centers. From visual inspection alone, the system could represent a probable pre-merger system or two independent clusters located at different cosmic distances that are projected close to each other on the plane of the sky.\\
\indent Figure\,\ref{Fig:4.4} also clearly reveals a wedge-like feature extending roughly orthogonal to the merging axis. The leading edge of the wedge is located about 280 kpc from the merging axis. Similar wedge-shaped structure has been reported in pre-merging cluster pairs such as 1E\,2216.0–0401 / 1E\,2215.7–0404 \citep{Gu_2019} and  1E\,2216.0-0401 / 1E\,2215.7-0404 \citep{Chen_2025}, where it is interpreted as shock feature. Confirming the presence of a shock requires a dedicated X-ray analysis. However, due to the limitations of the pipeline-processed X-ray data, we are currently unable to verify the existence of a shock at the wedge location. The overall X-ray distribution appears relatively regular, lacking strong asymmetries or extended disturbances, implying that the system is still dynamically young. Such a cluster morphology typically indicates an early-stage, pre-core-passage merger scenario like in the clusters Abell 851 and Abell 3528 \citep{schindler_1996,filippis_2003}. 
\subsubsection{Radio properties and ICM interaction}
\label{sec:4.1.2}
We found no evidence of cluster-wide diffuse radio halos or relics that could be associated with merger-driven shocks and turbulence, such as those seen in the systems like CIZA\,J2242.8+5301 (the ``Sausage'' relic; \citealt{weeren_2010}) and 1RXS\,J0603.3+4214 (the ``Toothbrush'' relic; \citealt{weeren_2012}), in either low or high-resolution radio maps. The large-scale morphology of the radio tail also appears to be regular and well-confined, suggesting that the surrounding ICM is not strongly turbulent or affected by large-scale sloshing.

To further test for the presence of any diffuse, cluster-scale radio emission, we subtracted compact sources in the visibility domain using a model obtained through high-resolution ($3''$) imaging and deconvolution. The field was then reimaged at a lower resolution to enhance the detectability of faint, extended structures (see Fig.~\ref{fig:4.2}). The $3''$ model resolution ensured that the diffuse radio tail was fully recovered, allowing a clean subtraction with minimal artefacts in the residual image. No emission attributable to a radio halo or inter-cluster bridge was detected down to a $3\sigma$ upper limit of $2.94\,\mathrm{mJy\,beam^{-1}}$ (based on an rms noise level of $0.98\,\mathrm{mJy\,beam^{-1}}$ in the residual map). Unlike well-studied pre-merger systems such as A399–A401, which hosts a clear radio bridge connecting the two clusters \citep{govoni_2019, pignataro_2024}, or A1758N–A1758S, where LOFAR observations revealed a candidate radio bridge along with double halos \citep{botteon_2020}, or post-merger systems like Abell\,2744 that exhibit a prominent giant radio halo \citep{pearce_2017}, we detected no comparable diffuse emission in our system. Our non-detection suggests that if a bridge or halo exists, its surface brightness must be below our sensitivity limit, consistent with a system in which merger-driven turbulence and shocks have not yet developed strongly enough to produce observable synchrotron emission. Also, note that in few cases, pre-merging clusters do not present large radio diffuse emissions \citep{Gu_2019,Lourenço_2023,Kurahara_2023}, which is consistent with our radio observations that no large scale diffuse radio emission is associated with either BCG.

Locally, while the peak of the X-ray emission marks the densest region of the ICM (i.e. the cluster core associated with the BCG), the radio galaxy is located well outside this core at a projected separation of \textbf{$\mathrm{R_{proj}\approx0.16}\,\mathrm{R_{200}}$} from the BCG, but clearly within the cluster region, and its orientation suggests motion roughly along the cluster’s large-scale gravitational potential gradient. The interaction between the radio galaxy and the ICM is further supported by the detailed radio morphology at intermediate and high angular resolution. In particular, the intermediate-resolution LOFAR image reveals a relatively abrupt termination of the radio tail, beyond which no diffuse synchrotron emission is detected. This location coincides spatially with a region where the X-ray surface brightness reduces significantly (refer to Fig.\,\ref{Fig:4.4}), indicating a decrease in the ambient ICM density. Such a correspondence suggests that the formation and visibility of the radio tail are closely tied to the galaxy’s encounter with the dense ICM \citep{begelman_1979,feretti_2012}.

In such scenarios, the radio jets are launched and subsequently bent by ram pressure as the host galaxy moves through the dense ICM \citep{Gunn_1972,jones_1979}. The observed narrow-angle morphology also supports a picture in which particle transport is dominated by bulk advection away from the nucleus, driven by the galaxy’s motion through the ICM \citep{Katz_1997,Hardcastle_2005}. This is clearly seen in the high-resolution image, where the southern jet seems to fully bend around at only $\sim1\,''$ of projected distance from the core, which would be $\mathrm{\sim5\,kpc}$ (projected). This is a really short distance within which the motion of the jet becomes dominated by its environment.

Taken together, the projected location of the radio galaxy within the cluster, the confinement and abrupt termination of the radio tail within dense X-ray emitting gas, the orientation along the plane of the sky, and the galaxy's morphology all point towards a scenario in which the galaxy has recently impinged through the ICM, and undergoing its first infall into the cluster environment. 
\subsection{Correlation between age and galaxy motion}\label{sec:4.2}
The velocity of a radio galaxy relative to the ICM is a critical parameter for constraining its dynamical state within the cluster. One commonly used approach is to relate the spatial extent of a radio tail to the radiative age of the synchrotron-emitting plasma, thereby deriving a characteristic velocity scale associated with the galaxy’s motion through the ICM. Assuming, based on the narrow-angle tailed morphology of the source, that the radio plasma in the tail is effectively at rest with respect to the ICM at sufficiently large distances from the AGN, a characteristic velocity scale can be approximated as
\begin{equation}
   \mathrm{v_{gal}=\frac{L_{proj}}{t_{spec}}} 
\end{equation}
 where $\mathrm{L_{proj}}$ is the projected length of the radio tail and $\mathrm{t_{spec}}$ is the spectral age. From our LOFAR and uGMRT imaging, we estimate the projected tail length to be approximately 300\,kpc, measured from the radio core to the end of the detectable emission. Using a best-fit spectral age of $\ 150\pm10$\,$\mathrm{Myr}$, this yields a projected velocity of $v_{gal}=1956 \pm 130$\,$\mathrm{km\,s^{-1}}$.

We emphasise that this estimate should be considered as a lower limit on the projected velocity component along the plane of the sky, rather than a direct measurement of the three-dimensional velocity. To place this velocity scale in the context of the cluster potential, we compared it with the host cluster's velocity dispersion. Since only the mass $M_{500}$ is available in the literature (refer to Table \ref{tab:cluster_properties} for the respective values),  we converted it to $M_{200}$ using the Colossus cosmology toolkit \citep[COsmology, haLO, and large-Scale StrUcture toolS;][]{diemer_2018}, a Python package designed for cosmological calculations. This yields
$M_{200} = 8.29\times 10^{14}\,M_\odot$, which we have used in the scaling
relation:
\begin{equation}
\displaystyle
    \frac{\sigma_{cl}}{\mathrm{km\,s^{-1}}} = A\left[\frac{h(z)M_{200}}{10^{15}M_{\odot}}\right]^\alpha
\end{equation}
as adopted in \citet{ferragamo_2021}, based on the scaling framework developed in \citet{evrard_2008}. This gives a cluster velocity dispersion of $\sigma_{cl}=1186 \pm 86\, \mathrm{km\,s^{-1}}$. The estimated velocity of the galaxy is therefore significantly above the average internal motions within the cluster by a factor of roughly 1.7. Furthermore, for the inferred cluster's mass, the escape velocity within $\mathrm{R_{500}}$ is $\mathrm{v_{esc}\approx2113\,km\,s^{-1}}$  \citep[within systematic uncertainties;][]{Pratt_2009}. This suggests that even though the galaxy may still be gravitationally bound to the cluster, it is moving at substantial fraction of the escape velocity, of order of $\mathrm{\sim 0.9\,v_{esc}}$.

Additional constraints on the galaxy’s motion were provided by photometric redshift measurements from  the Dark Energy Spectroscopic Instrument Data Release 9 \citep[DESI DR9;][]{Li_2023}, which suggested a line-of-sight velocity of $\mathrm{v_{los}\approx3069\,\pm\,2146\,km\,s^{-1}}$ relative to the cluster's redshift. When combined in quadrature with the projected velocity inferred from the radio tail, this implies a characteristic three-dimensional velocity of $\mathrm{v_{3D}\approx3640\pm1811\,km\,s^{-1}}$. Despite the large uncertainties associated with the photometric redshift–based line-of-sight velocity, the current data are most consistent with a velocity that is substantially in excess of the cluster's escape velocity. This strongly supports a scenario in which the galaxy is not a virialized cluster member, but a more recently accreted system that  is undergoing its first interaction with the ICM moving along a trajectory consistent with the first infall.
\begin{figure}
    \centering
    \includegraphics[width=\linewidth]{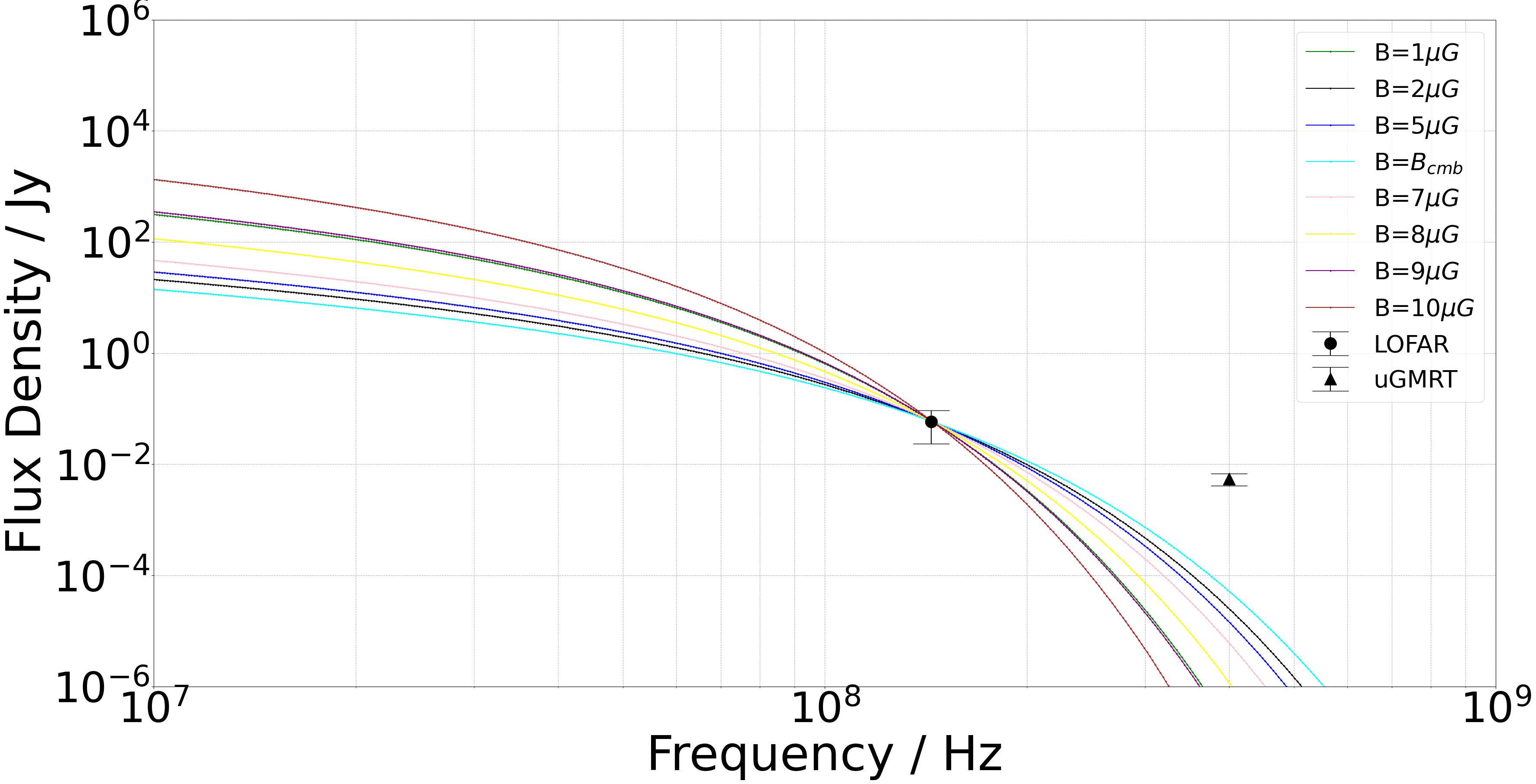} 
    \caption{Spectral ageing model at a dynamical age of 270 Myr, showing the flux density as a function of frequency for different magnetic field strengths (B). The black points indicate the observed flux densities for comparison.}
    \label{Fig:4.3}
\end{figure}

\subsection{Consistency of Spectral Ages with Dynamical Timescales}\label{sec:4.3}
To test the robustness of the conclusions drawn in Section~\ref{sec:4.2}, we explored the inverse scenario: assuming that the tailed radio galaxy was a virialized cluster member moving with the cluster velocity dispersion, we examined whether the observed spectral indices could be reconciled with this dynamical condition through standard spectral ageing models. If the galaxy moved at the typical cluster velocity ($\sigma_v = 1186~\mathrm{km\,s^{-1}}$), the corresponding dynamical age of the radio tail, defined as the travel time by the electrons from the AGN to the tip of the observed tail, would be 
\begin{equation}
  T_{\rm dyn} = \frac{324~\mathrm{kpc}}{1186~\mathrm{km\,s^{-1}}} \approx 270~\mathrm{Myr}.  
\end{equation}
This represented the characteristic timescale expected if the host galaxy had long been bound to the cluster potential.

Figure~\ref{Fig:4.3} showed the predicted spectral ageing curves at this fixed dynamical age of $270~\mathrm{Myr}$ for a range of magnetic field strengths ($B = 1$–$10~\mathrm{\mu G}$), including the CMB-equivalent field ($B_{\rm CMB} = 3.25(1+z)^2~\mathrm{\mu G}$). Under these assumptions, the modeled spectra fell significantly below the observed uGMRT flux density at $400\,\mathrm{MHz}$ when normalized to the LOFAR flux at $144\,\mathrm{MHz}$. This systematic shortfall demonstrated that, if the galaxy moved with the cluster’s velocity dispersion, standard ageing models could not reproduce the observed spectral curvature. The inconsistency implied that the effective radiative age of the electrons was significantly shorter than the $\sim270~\mathrm{Myr}$ dynamical timescale expected for a virialized member, requiring a higher velocity consistent with the $\sim2000~\mathrm{km\,s^{-1}}$ inferred in Section~\ref{sec:4.2}. 

In our analysis, we have used resolved spectral index maps and spatially localized tail regions to mitigate one of the main biases in integrated spectral studies, the mixing of electron populations with different radiative ages within unresolved regions (\citealt{harwood_2016}; \citealt{turner_2018}). The persistence of the mismatch even in this resolved framework strengthened the interpretation that the host galaxy is not moving at the cluster's velocity dispersion, but is instead an infalling system.

Lastly, the only key uncertainty in this analysis lies in the assumed magnetic field strength, which entered non-linearly into the spectral age estimates. In the absence of direct constraints, even moderate deviations from the adopted $B$ produced substantial changes in the derived ages, making a factor-of-two discrepancy between the spectral and dynamical timescales unsurprising. Moreover, standard spectral ageing models assumed a single injection episode and a spatially uniform magnetic field, both of which were idealizations. Deviations from these assumptions, such as reacceleration, intermittent jet activity, or magnetic field inhomogeneities, could further influence the inferred ages. However, within reasonable limits of these uncertainties, no combination of magnetic field strength, injection index, or spectral age could simultaneously reproduce the observed spectral indices and a dynamical age 270 Myr.
\section{Conclusions}\label{sec:conclusion}

In this study, we presented the first focused radio analysis of the galaxy cluster MACS\,J1354.6+7715, utilizing high-resolution imaging from LOFAR and uGMRT. Our observations revealed a narrow-angle tailed (NAT) radio galaxy embedded within the cluster, whose morphology and spectral properties provide valuable insight into the dynamical state of the system. By constructing resolved spectral index maps of the radio tail and incorporating multi-wavelength data from Chandra X-ray and HST optical observations, we examined the interplay between the AGN, its host galaxy, and the surrounding ICM. Our key findings are summarized below:
\begin{itemize}
    \item The radio galaxy exhibits a well-defined, extended radio tail indicating an ongoing interaction with the ICM, likely driven by ram pressure stripping. The radio jet morphology suggests that environmental interactions are significantly influencing the evolution of the radio galaxy, impacting its trajectory and jet structure.
    \item Spectral index maps constructed from LOFAR and uGMRT data reveal a smooth steepening along the tail, consistent with synchrotron ageing of the electron population. This gradient suggests continuous plasma injection with minimal reacceleration or mixing.
    \item The absence of morphological features typical of radio relics in the high-resolution radio image, together with the lack of detectable diffuse halo or inter-cluster bridge emission in the compact source subtracted image, indicates that no strong merger-driven shocks or large-scale turbulence are currently present. When combined with the bimodal but relatively undisturbed X-ray surface brightness distribution and the coherent, gently bent morphology of the tailed radio galaxy, we conclude that MACS\,J1354.6+7715 is a pre-merger candidate, with two subclusters likely approaching but not yet having undergone core passage.
    \item The estimated velocity of the host galaxy ($v_{\mathrm{gal}} = 1956 \pm 130\,\mathrm{ km\,s^{-1}}$), derived from the projected length of the radio tail ($\sim$\nobreakspace300~kpc) and its spectral age ($150\pm10$\,Myr), is of order $\mathrm{\sim 0.9\,v_{\rm esc}}$ and significantly exceeds the cluster’s velocity dispersion ($\sigma_{\mathrm{cl}} = 1186 \pm 86\,\mathrm{ km\,s^{-1}}$), suggesting that the galaxy is a recent infaller rather than a virialized cluster member.
    \item A comparison between the spectral and dynamical ages revealed a significant mismatch: for the dynamical age expected if the galaxy were a cluster member ($\sim270~\mathrm{Myr}$), standard spectral ageing models underpredicted the observed flux density at higher frequencies, even when using spatially resolved spectral index maps. This discrepancy suggests that the galaxy is moving faster than the cluster dispersion, although uncertainties in the assumed magnetic field strength, potential reacceleration, and simplifications in the models (e.g., uniform $B$ and a single injection episode) prevent precise determination of the absolute infalling speed.
\end{itemize}
These findings highlight the critical role of environmental effects in shaping the evolution of galaxies within clusters. The interaction between the infalling galaxy and the dense ICM alters its radio morphology and provides a valuable probe of the cluster’s dynamical history. The presence of such an extreme case highlights the importance of high-resolution, multi-frequency radio imaging in tracing AGN activity and understanding galaxy evolution in dense environments. It also underscores the need for deeper, multi-wavelength surveys to identify similar infall scenarios in other merging clusters.

Despite these insights, several open questions remain, particularly the persistent discrepancy between spectral and dynamical age estimates, even in spatially resolved analyses. While the combined LOFAR and uGMRT data enabled detailed morphological and spectral study, upcoming facilities such as LOFAR 2.0 and the SKA will provide enhanced sensitivity and broader frequency coverage. This will allow for more precise measurements of spectral curvature and break frequencies, reducing dependence on assumptions about magnetic field strength and electron injection history, and improving the accuracy of spectral age estimates.\\

\section*{Acknowledgements}
This research made use of data from the LOFAR Two-metre Sky Survey \citep[LoTSS;][]{Shimwell_2017,Shimwell_2022}. LOFAR \citep{vanhaarleem_2013} is the Low Frequency Array designed and constructed by ASTRON and has facilities in several countries that are owned by various parties (each with their own funding sources) and collectively operated by the International LOFAR Telescope (ILT) foundation under a joint scientific policy. We thank the staff of the GMRT for making these observations possible. The GMRT is run by the National Centre for Radio Astrophysics of the Tata Institute of Fundamental Research (NCRA–TIFR). We also thank the developers of \textsc{LINC}, \textsc{WSClean}, \textsc{DP3}, and \textsc{BRATS}, which were used for calibration, imaging, and spectral modelling. LKM is grateful for support from a UKRI Future Leaders Fellowship [MR/Y020405/1]. AG and LKM acknowledge support from the International Science Partnerships Fund (ISPF) through STFC grant ST/Y004159/1. RT is grateful for support from the UKRI Future Leaders Fellowship (grant MR/T042842/1). This work was supported by the STFC [grants ST/T000244/1, ST/V002406/1]. R.K. and A. P acknowledge the support of the Department of Atomic Energy, Government of India, under project no. 12-R\&D-TFR-5.02-0700.

\section*{Data Availability}
The LOFAR data used in this paper were accessed from the $\mathrm{LOFAR}$ Long Term Archive (LTA) under project code \textsc{LT14\_004}.  
The LoTSS\footnote{\url{https://lofar-surveys.org}} data products are publicly available online. The uGMRT data used in the paper are available in the GMRT online archive\footnote{\url{https://naps.ncra.tifr.res.in/goa/}} under project code \textsc{37\_128}.
All derived data products and imaging scripts will be shared upon reasonable request to the corresponding author.



\bibliographystyle{mnras}
\bibliography{example} 








\bsp	
\label{lastpage}
\end{document}